\documentclass[twocolumn,10pt]{IEEEtran}
\usepackage{epstopdf}
\usepackage{algorithm,algorithmic,amsbsy,amsmath,amssymb,epsfig,bbm,mathrsfs,fancyhdr,fancyvrb,subfigure,xcolor}

\newcommand{\beq}{\begin{equation}}
\newcommand{\eeq}{\end{equation}}
\newcommand{\beqn}{\begin{eqnarray}}
\newcommand{\eeqn}{\end{eqnarray}}

\newcommand{\qed}{\nobreak \ifvmode \relax \else
      \ifdim\lastskip<1.5em \hskip-\lastskip
      \hskip1.5em plus0em minus0.5em \fi \nobreak
      \vrule height0.75em width0.5em depth0.25em\fi}

\begin{document}

\title{Evolution Towards 5G Multi-tier Cellular Wireless Networks: An Interference Management Perspective}
\author{Ekram Hossain,  Mehdi Rasti, Hina Tabassum, and Amr Abdelnasser\thanks{E. Hossain, H. Tabassum, and A. Abdelnasser are with the Department of Electrical and Computer Engineering at the University of Manitoba, Canada. M. Rasti is with the Department of Computer Engineering and Information Technology, 
Amirkabir University of Technology, Iran. This work was supported  by the Natural Sciences and Engineering Research Council of Canada (NSERC) Strategic  Project Grant STPGP 430285-12.}}

\maketitle

\begin{abstract}
The evolving fifth generation (5G) cellular wireless networks are envisioned to overcome the fundamental challenges of existing cellular networks, e.g.,  higher data rates, excellent end-to-end performance and user-coverage in hot-spots and crowded areas with lower latency, energy consumption and cost per information transfer. To address these challenges, 5G systems will  adopt a multi-tier architecture consisting of macrocells, different types of licensed small cells, relays, and device-to-device (D2D) networks to serve users with different quality-of-service (QoS) requirements in a spectrum and energy-efficient manner.  Starting with the visions and requirements of  5G multi-tier networks, this article outlines the challenges of interference management (e.g., power control, cell association) in these networks with shared spectrum access (i.e., when the different network tiers share the same licensed spectrum). 
It is argued that the existing  interference management schemes  will not be able to address the interference management problem in prioritized 5G multi-tier networks where users in different tiers have different priorities for channel access.
In this context, a  survey and qualitative comparison of the existing cell association and power control schemes is  provided to demonstrate their limitations for interference management in 5G networks.
Open challenges are highlighted and guidelines are provided to modify the existing schemes in order to overcome these limitations and make them suitable for the emerging 5G systems.

\end{abstract}

\begin{keywords}
5G cellular wireless, multi-tier networks, interference management, cell association, power control.
\end{keywords}

\IEEEpeerreviewmaketitle
\section{Introduction}

To satisfy the ever-increasing demand for mobile broadband communications, the IMT-Advanced (IMT-A) standards have been ratified by the International Telecommunications Union (ITU) in November 2010 and the fourth generation (4G) wireless communication systems are currently being deployed worldwide. The standardization for LTE Rel-12, also known as LTE-B, is also ongoing and expected to be finalized in 2014. Nonetheless, existing wireless systems will not be able to deal with the thousand-fold increase in total mobile broadband data  \cite{erxson}  contributed by new applications and services such as  pervasive 3D multimedia, HDTV, VoIP, gaming, e-Health, and Car2x communication. In this context, the fifth generation (5G) wireless communication technologies are expected  to attain 
1000 times higher mobile data volume per unit area, 10-100 times higher number of connecting devices and user data rate, 10 times longer battery life and 5 times reduced latency \cite{metis}. While for 4G networks the single-user average data rate is expected to be 1 Gbps,  it is postulated that cell data rate of the order of 10~Gbps will be a key attribute of 5G networks.  

5G wireless networks are expected to be a mixture of network tiers of different sizes, transmit powers, backhaul connections, different radio access technologies (RATs) that are accessed by an unprecedented numbers of smart and heterogeneous wireless devices.  This architectural enhancement along with the advanced physical communications technology such as high-order spatial multiplexing multiple-input multiple-output (MIMO) communications  will  provide higher aggregate capacity for more simultaneous users, or higher level spectral efficiency, when compared to the 4G networks. Radio resource and interference management will be a key research challenge in multi-tier and heterogeneous 5G cellular networks. The traditional methods for radio resource and interference management (e.g., channel allocation, power control, cell association or load balancing) in single-tier networks (even some of those developed for two-tier networks) may not be efficient in this environment and a new look into the interference management problem will be required. 
	
First, the article outlines the visions and requirements of  5G cellular wireless systems.  Major research challenges are then highlighted from the perspective of interference management  
when the different network tiers share the same radio spectrum. A comparative analysis of the existing approaches for distributed cell association and power control (CAPC) is then provided followed by a discussion on their limitations  for  5G multi-tier cellular networks. Finally, a number of suggestions are provided to modify the existing CAPC schemes to overcome these limitations.

\section{Visions and Requirements for 5G Multi-Tier Cellular Networks}

5G mobile and wireless communication systems will require a mix of new system concepts to boost the spectral and energy efficiency. 
The visions and requirements  for 5G wireless systems are outlined below.

\begin{itemize}

\item {\em Data rate and latency}: 
For dense urban areas, 5G networks are envisioned to  enable an experienced data rate of 300~Mbps and 60~Mbps in downlink and uplink, respectively,  in 95\% of locations and time \cite{metis}. The end-to-end latencies are expected to be in the order of 2 to 5 milliseconds. The detailed requirements for different scenarios are listed in \cite{metis}.

\item {\em Machine-type Communication (MTC) devices}: The number of traditional human-centric wireless devices with Internet connectivity (e.g., smart phones, super-phones, tablets) may be outnumbered by  MTC devices which can be used in vehicles, home appliances, surveillance devices, and sensors.

\item {\em Millimeter-wave communication}:  To satisfy the exponential increase in traffic and the addition of different devices and services,  additional spectrum beyond what was previously allocated to 4G standard is sought for.   The use of millimeter-wave frequency bands (e.g., 28~GHz and 38~GHz bands) is a potential candidate to overcome the problem of scarce spectrum resources since it allows transmission at wider bandwidths than conventional 20 MHz channels for 4G systems. 

\item {\em Multiple RATs}:  
5G is not about replacing the existing technologies, but it is about enhancing and supporting them with new technologies  \cite{erxson}. In 5G systems, the  existing RATs, including GSM (Global System for Mobile Communications), HSPA$+$ (Evolved High-Speed Packet Access), and LTE, will continue to evolve 
to provide a superior system performance. They will also be accompanied by some new technologies (e.g., beyond LTE-Advanced). 

\item {\em Base station (BS) densification}: 
BS densification is an effective methodology to meet the requirements of 5G wireless networks. Specifically, in 5G networks, there will be deployments of a large number of low power nodes, relays, and device-to-device (D2D) communication links with much higher density than today's macrocell networks. Fig. \ref{Multi-Tier} shows such a multi-tier network with a macrocell overlaid by relays, picocells, femtocells, and D2D links.  The adoption of multiple tiers in the cellular network architecture  will result in better performance in terms of capacity, coverage, spectral efficiency, and total power consumption, provided that the inter-tier and intra-tier interferences are well managed.

\begin{figure*}[ht]
\begin{center}
\includegraphics[width=7 in]{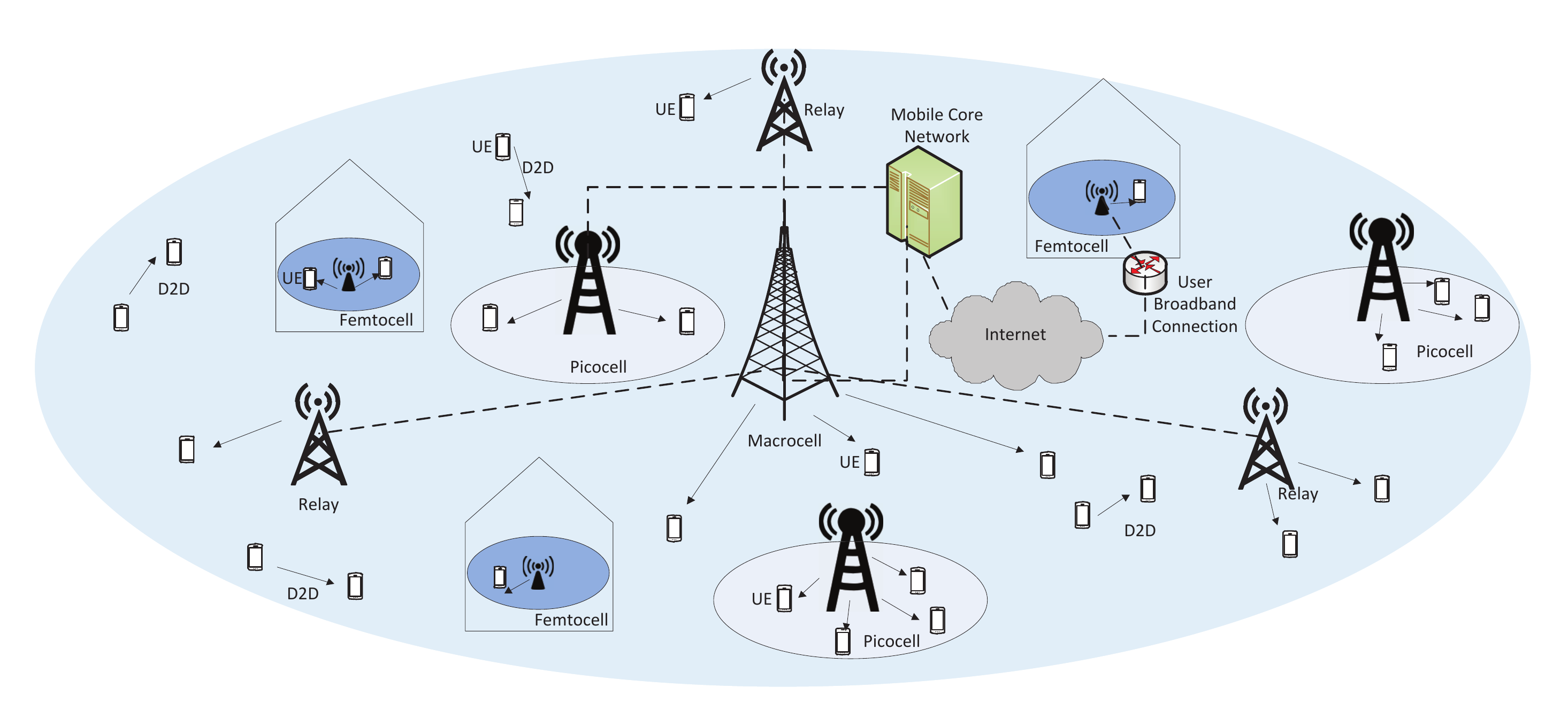}
\caption [c]{A multi-tier network composed of macrocells, picocells, femtocells, relays, and D2D links. Arrows indicate wireless links, whereas the dashed lines denote the  backhaul connections.}
\label{Multi-Tier}
\end{center}
\end{figure*}
	
\item {\em Prioritized spectrum access}:  The notions of both traffic-based and tier-based priorities will exist in 5G networks. Traffic-based priority arises from the different requirements of the users (e.g., reliability and latency requirements, energy constraints),  whereas the tier-based priority is for users belonging to different network tiers.  For example, with shared spectrum access among macrocells and femtocells in a two-tier network, femtocells create ``dead zones" around them in the downlink  for macro users. 
Protection should, thus, be guaranteed for the macro users. Consequently, the macro and femtousers  play the role of high-priority users (HPUEs) and low-priority users (LPUEs), respectively. In the uplink direction, the macrocell users at the cell edge typically transmit with high powers which generates high uplink interference to  nearby  femtocells. Therefore, in this case, the user priorities should get reversed. Another example is a D2D  transmission where different devices  may opportunistically access the spectrum to establish a communication link between them
provided that the interference introduced to the cellular users  remains below a given threshold. In this case, the D2D users play the role of LPUEs whereas the cellular users play the role of HPUEs. 

\item {\em Network-assisted D2D communication}:
In the LTE Rel-12 and beyond, focus will be on network controlled D2D communications, where the macrocell BS performs control signaling in terms of synchronization, beacon signal configuration and providing identity and security management \cite{Astely2013}. This feature will extend in 5G  networks to allow other nodes, rather than the macrocell BS, to have the control. For example, consider a D2D link at the cell edge and the direct link between the D2D transmitter UE to the macrocell is in deep fade, then the relay node can be responsible for the control signaling of the D2D link (i.e., relay-aided D2D communication).

\item {\em Energy harvesting for energy-efficient communication}: 
One of the main challenges in 5G wireless networks is to improve the energy efficiency of the  battery-constrained wireless devices. To prolong the battery lifetime as well as to improve the energy efficiency, an appealing solution is to harvest energy  from environmental energy sources (e.g., solar and wind energy). Also, energy can be harvested from ambient radio signals (i.e., RF energy harvesting) with reasonable efficiency over small distances. The havested energy could be used for D2D communication or communication within a small cell. In this context, simultaneous wireless information and power transfer (SWIPT) is a promising technology for 5G wireless networks. However, practical circuits for harvesting energy are not yet available since the conventional receiver architecture is designed for information transfer only and, thus, may not be optimal for SWIPT. This is due to the fact that both information and power transfer  operate with different power
sensitivities at the receiver (e.g., -10dBm and -60dBm for energy and information receivers, respectively) \cite{zhou2012wireless}. Also, due to the potentially low efficiency of energy harvesting from ambient radio signals, a combination of different energy harvesting technologies may be required for macrocell communication. 

\end{itemize}

\section{Interference Management Challenges in 5G Multi-tier Networks}

The key challenges for interference management in 5G multi-tier networks will arise due to the following reasons which affect the interference dynamics in the uplink and downlink of the network: (i)  heterogeneity and dense deployment of wireless devices, 
(ii) coverage and traffic load imbalance due to varying transmit powers of different BSs in the downlink, (iii) public or private access restrictions in different tiers that lead to diverse interference levels, and 
(iv) the priorities in accessing channels of different frequencies and resource allocation strategies. Moreover,  the introduction of carrier aggregation, cooperation among BSs (e.g., by using coordinated multi-point transmission (CoMP)) as well as direct communication among users (e.g., D2D communication) may further complicate the dynamics of the interference.  
The above factors translate into the following key challenges.

\begin{itemize}

\item {\em Designing optimized cell association and power control (CAPC) methods for multi-tier networks}:
Optimizing the cell associations and transmit powers of users in the uplink or the transmit powers of BSs  in the downlink are  classical techniques to simultaneously enhance the system performance in various aspects such as interference mitigation, throughput maximization, and reduction in power consumption. Typically, the former is needed to maximize spectral efficiency, whereas the latter is required to minimize the power  (and hence minimize the interference to other links) while keeping the desired link quality. 
Since it is not efficient to connect to a congested BS despite its high achieved signal-to-interference ratio (SIR), cell association should also consider the status of each BS (load) and the channel state of each UE.
The increase in the number of available BSs along with multi-point transmissions and carrier aggregation provide multiple degrees of freedom for resource allocation and cell-selection strategies. For power control, the priority of different tiers need also be maintained by incorporating the quality constraints of HPUEs.

Unlike downlink, the transmission power in the uplink depends on the user's battery power irrespective of the type of BS with which users are connected. The battery power does not vary significantly from user to user; therefore,  the problems of coverage and traffic load imbalance may not exist in the uplink.  This leads to considerable asymmetries between the uplink  and downlink user association policies. Consequently, the optimal solutions for downlink CAPC problems may not be optimal for the uplink. It is therefore  necessary to develop joint optimization frameworks that can provide near-optimal, if not optimal, solutions for both uplink and downlink. Moreover, to deal with this issue of asymmetry, separate uplink and downlink optimal solutions are also useful as far as mobile users can connect with two different BSs for uplink and downlink transmissions which is expected to be the case in 5G multi-tier cellular networks \cite{Astely2013}.

\item {\em Designing efficient methods to support simultaneous association to multiple BSs}:
Compared to existing CAPC schemes in which each user can associate to a single BS, simultaneous connectivity to several BSs could be possible in 5G multi-tier network. This would enhance the system throughput and reduce the outage ratio by effectively utilizing the available resources, particularly for cell edge users. Thus the existing CAPC schemes should be extended to efficiently support simultaneous association of a user to multiple BSs and determine under which conditions a given UE is associated to which BSs in the uplink and/or downlink.

\item{\em Designing efficient methods for cooperation and coordination among multiple tiers}:
Cooperation and coordination among different tiers will be a key requirement to mitigate interference in 5G networks. Cooperation between the macrocell and small cells was proposed for LTE Rel-12 in the context of soft cell, where the UEs are allowed to have dual connectivity by simultaneously connecting to the macrocell and the small cell for  uplink and downlink communications or vice versa \cite{Astely2013}. As has been mentioned before in the context of asymmetry of transmission power in uplink and downlink,  a UE may experience the highest downlink power  transmission from the macrocell, whereas the highest uplink path gain may be from a nearby small cell. In this case, the UE can  associate to the macrocell in the downlink and to the small cell in the uplink. CoMP schemes based on cooperation among BSs in different tiers (e.g., cooperation between macrocells and small cells) can be developed to mitigate interference in the network.  Such schemes need to be adaptive and consider user locations as well as channel conditions to maximize the spectral and energy efficiency of the network.
This cooperation however, requires tight integration of low power nodes into the network through the use of reliable, fast and low latency backhaul connections which will be a major technical issue for upcoming multi-tier 5G networks.

\end{itemize}

In the remaining of this article, we will focus on the review of existing  power control and cell association strategies to demonstrate their limitations for interference management in 5G multi-tier prioritized cellular networks (i.e., where users in different tiers have different priorities depending on the location, application requirements and so on). Design guidelines will then be provided to overcome these limitations. Note that issues such as channel scheduling in frequency domain, time-domain interference coordination techniques (e.g., based on almost blank subframes),  coordinated multi-point transmission, and spatial domain techniques (e.g., based on smart antenna techniques) are not considered in this article.

\section {Distributed Cell Association and Power Control Schemes: Current State of the Art}

\subsection{Distributed  Cell Association Schemes}
The  state-of-the-art cell association schemes that are currently under investigation for  multi-tier cellular networks are reviewed and their limitations are explained below.
\begin{itemize}
\item {\em Reference Signal Received Power (RSRP)-based scheme} \cite{sangiamwong2011investigation}: A user is  associated with the BS whose signal is received with the largest average strength. A  variant of RSRP, i.e., Reference Signal Received Quality (RSRQ) is also used for cell selection in LTE single-tier networks which is similar to the signal-to-interference (SIR)-based cell selection where a user selects a BS communicating with which gives the highest SIR.  In single-tier networks with uniform traffic, such a criterion may maximize the network throughput. However, due to varying transmit powers  of different BSs in  the downlink of multi-tier networks, such cell association policies can create a huge traffic load imbalance.
This phenomenon leads to overloading of high power tiers while leaving  low power tiers underutilized.  

\item {\em Bias-based Cell Range Expansion (CRE)} \cite{guvenc2011capacity}: The idea of 
CRE has been emerged as a remedy to the problem of load imbalance in the downlink. It aims to increase the downlink coverage footprint of low power BSs by adding a positive bias to their signal strengths (i.e., RSRP or RSRQ). Such BSs are referred to as \emph{biased BSs}. This biasing allows more users to associate with low power or biased BSs and thereby achieve a better  cell load balancing. Nevertheless, such off-loaded users may experience  unfavorable channel from the biased BSs and strong interference from the unbiased high-power BSs. {The trade-off between cell load balancing and  system throughput therefore strictly depends on the selected bias values which need to be optimized in order to maximize the system utility.} In this context, a baseline approach in LTE-Advanced is to ``orthogonalize" the transmissions of the biased and unbiased  BSs in time/frequency domain such that an interference-free zone is created.

\item {\em Association based on Almost Blank Sub-frame (ABS) ratio} \cite{absratio}:
The ABS technique uses time domain orthogonalization in which specific sub-frames are left blank by the unbiased BS and off-loaded users  are scheduled within these sub-frames to avoid inter-tier interference. This improves the overall throughput of the off-loaded users by sacrificing the time sub-frames and  throughput of the unbiased BS. The larger bias values result in higher degree of offloading and thus require more blank sub-frames to protect the offloaded users.  
Given a specific number of ABSs or  the \emph{ratio of blank over total number of sub-frames (i.e., ABS ratio)} that ensures the minimum throughput of the unbiased BSs, this criterion allows a user to select a cell with maximum ABS ratio and may  even associate with the unbiased BS if ABS ratio decreases significantly.

\end{itemize}

A qualitative comparison among these cell association schemes is given in Table~\ref{compare:association}. The specific key terms used in Table~\ref{compare:association} are defined as follows:
{\em channel-aware} schemes depend on the knowledge of instantaneous channel and transmit power at the receiver. The {\em interference-aware} schemes depend on the knowledge of instantaneous interference at the receiver. The {\em load-aware} schemes depend on the  traffic load information (e.g., number of users).  The {\em resource-aware} schemes require the resource allocation information (i.e., the chance of getting a channel or the proportion of resources available in a cell). The    {\em priority-aware} schemes require the information regarding the priority of different tiers and allow a protection to HPUEs.

\begin{table*}
\caption{Qualitative comparison of existing cell association schemes for  multi-tier networks} 
\centering 
\begin{tabular}
{|p{2.4cm}|p{2.95cm}| p{2.75cm}| p{2.55cm}|p{2.55cm}|p{2.75cm}|p{2.75cm}|p{2.55cm}|} 
\hline\hline 
& RSRP \cite{sangiamwong2011investigation}
& RSRQ \cite{sangiamwong2011investigation}
& CRE \cite{guvenc2011capacity}
& ABS ratio \cite{absratio}
\\
\hline\hline
Objective
&	Maximize  received signal power
&	Maximize SIR 
&  Balance traffic load 
&  Maximize rate and balance traffic load
\\
\hline
{Applicability} 
& Uplink and downlink 
& Uplink and downlink 
& Downlink 
& Downlink 
\\\hline
{Channel-aware} 
&	\checkmark 
& \checkmark 
& \checkmark 
& \checkmark  
\\
\hline
{Interference-aware} 
& \text{\sffamily X}	
& \checkmark 
& \checkmark  
& \checkmark 
\\
\hline 
{Traffic load-aware} 
& \text{\sffamily X}	
& \text{\sffamily X} 
& \checkmark 
& \checkmark 
\\
\hline
Resource-aware 
& \text{\sffamily X} 
& \text{\sffamily X} 
& \text{\sffamily X} 
& \checkmark
\\ 
\hline
Priority-aware 
&\text{\sffamily X} 
& \text{\sffamily X} 
& \text{\sffamily X} 
& \checkmark
\\ 
\hline\hline
\end{tabular}
\label{compare:association}
\end{table*}

All of the above mentioned schemes are independent, distributed, and can be incorporated with any type of power control scheme. Although simple and tractable,  the standard cell association schemes, i.e., RSRP, RSRQ, and CRE are unable to guarantee the optimum  performance in multi-tier networks unless critical parameters, such as bias values, transmit power of the users in the uplink and BSs in the downlink, resource partitioning, etc.  are optimized.  

\subsection{Distributed Power Control Schemes}
From a user's point of view, the objective of power control  is to support a user with its minimum acceptable throughput, whereas from a system's point of view it is to maximize the aggregate throughput. In the former case, it is required to compensate for the near-far effect by allocating higher power levels to users with poor channels as compared to UEs with good channels. In the latter case, high power levels are allocated to users with best channels and very low (even zero) power levels are allocated to others. The {\em aggregate transmit power, the outage ratio,} and the {\em aggregate throughput  (i.e., the sum of achievable rates by the UEs)} are the most important measures to compare the performance of different power control schemes. The outage ratio of a particular tier  can  be expressed as the ratio of the number of UEs supported by a tier with their minimum target  SIRs and the total number of UEs in that tier. 

Numerous power control schemes have been proposed in the literature for single-tier cellular wireless networks. According to the corresponding objective functions and assumptions, the schemes can be classified into the following four types. 

\begin{itemize}
\item {\em Target-SIR-tracking  power control (TPC)} \cite{fos93}: In the TPC, each UE  tracks its own predefined fixed target-SIR. The TPC  enables the UEs to achieve their fixed target-SIRs at minimal aggregate transmit power, assuming that the target-SIRs are feasible. However,  when the system is infeasible, all non-supported UEs (those who cannot obtain their target-SIRs) transmit at their maximum power, which causes unnecessary power consumption and  interference to other users, and therefore, increases the number of non-supported UEs. 

\item {\em TPC with gradual removal  (TPC-GR)}~\cite{rasti11}, \cite{rastisoft}, and \cite{bergg01}:
To decrease the outage ratio of the TPC in an infeasible system, a number of TPC-GR algorithms were proposed  in which non-supported users reduce their transmit power \cite{rastisoft} or are gradually removed  \cite{rasti11,bergg01}.
		    
\item {\em Opportunistic power control (OPC)}~\cite{leung06}:
 From the system's point of view, OPC allocates high power levels to users with good channels (experiencing high path-gains and low interference levels) and very low power to users with poor channels. In this algorithm, a small difference in path-gains between two users may lead to a large difference in their actual throughputs \cite{leung06}. OPC improves the system performance at the cost of reduced fairness among users.

\item {\em Dynamic-SIR tracking power control (DTPC)}~\cite{rastidtpc}:
When the target-SIR requirements for users are feasible, TPC causes users to exactly hit their fixed target-SIRs  even if additional resources are still available that can otherwise be used to achieve higher SIRs (and thus better throughputs).  Besides, the fixed-target-SIR assignment is suitable only for voice service for which reaching a SIR value higher than the given target value does not  affect the service quality significantly. In contrast, for data services, a higher SIR results in a better throughput, which is desirable. 
The DTPC algorithm was proposed in \cite{rastidtpc} to address the problem of system throughput maximization subject to a given feasible lower bound for the achieved SIRs of all users in  cellular networks. In DTPC, each user dynamically sets its target-SIR by using TPC and OPC in a selective manner. It was shown that when the minimum acceptable target-SIRs are feasible, the actual SIRs received by some users can be dynamically increased  (to a value higher than their minimum acceptable target-SIRs)  in a distributed manner  so far as the required resources are available and the system remains feasible (meaning that reaching the minimum target-SIRs for the remaining users are guaranteed). This enhances the system throughput (at the cost of higher power consumption) as compared to TPC.

\end{itemize}

The aforementioned state-of-the-art distributed power control schemes for satisfying various objectives in single-tier wireless cellular networks are unable to address the interference management problem in prioritized 5G multi-tier  networks. This is due to the fact that they do not guarantee that the total interference caused  by the LPUEs to the HPUEs  remain within tolerable limits, which can lead to the SIR outage of some HPUEs. 
Thus there is a need to modify the existing schemes such that LPUEs track their objectives while limiting their transmit power to maintain a given interference threshold at HPUEs. 

A qualitative comparison among various state-of-the-art power control problems with different objectives and constraints and their corresponding existing distributed solutions are shown in Table~II.  This table also shows how these schemes can be modified  and generalized for designing CAPC schemes for prioritized 5G multi-tier networks.
 
\begin{table*}[h]
	\caption{Comparison among various power control optimization problems and their corresponding distributed solutions  for single-tier networks, along with their application and generalization for designing CAPC schemes in 5G networks} 
	\centering 
	\begin{tabular}
{|p{3.5cm}|p{3.5cm}| p{3.5cm}| p{3.5cm}|} 
	\hline\hline  
	& Power control in single-tier networks & Power control in prioritized multi-tier networks & Joint cell association and power control in multi-tier networks\\
	\hline\hline
\raisebox{0.25ex}{Optimization problem (P1)} 
	& With fixed BS assignment, minimize aggregate power subject to  minimum target-SIR for all users 
	& With fixed BS assignment, minimize aggregate power subject to  minimum (different) target-SIRs for users in different tiers
	& Minimize aggregate power subject to  minimum (different) target-SIRs for users in different tiers and obtain BS assignment
\\
	\hline
	Distributed solutions
	&TPC \cite{fos93}
	&TPC \cite{fos93}
	& Minimum effective interference based cell association+TPC \cite{yates1995integrated}
\\
	\hline\hline
\raisebox{0.25ex}{Optimization problem (P2)} 
	& Minimize outage ratio of all users
	& Minimize outage ratio of LPUEs subject to zero-outage for HPUEs 
	&  Minimize outage ratio of all users in different tiers and obtain BS assignment\\
	\hline
	Distributed solutions
	&TPC-GR \cite{bergg01}, \cite{rasti11}, \cite{rastisoft}
	&Open problem
	&Open problem
\\
	\hline\hline
\raisebox{0.25ex}{Optimization problem (P3)}  
	& Maximize aggregate throughput of all users
	& Maximize aggregate throughput of all users subject to zero-outage for HPUEs  
	& Maximize aggregate throughput of all users and obtain BS assignment
\\
	\hline
	Distributed solutions &OPC \cite{leung06}
	&Open problem
   &Open problem
	\\
	\hline\hline
\raisebox{0.25ex}{Optimization problem (P4)} 
	& Maximize aggregate throughput of all users subject to  minimum target-SIR for all users
	&Maximize aggregate throughput of LPUEs subject to  minimum target-SIR for all users
   & Maximize aggregate throughput of all users subject to  minimum target-SIR for all users and obtain BS assignment
\\
	\hline
	Distributed solutions
	& DTPC \cite{rastidtpc}
	&Open problem
	&Minimum effective-interference-based cell association + HPC \cite{vudistributed}
\\
	\hline\hline
	\end{tabular}
	\label{tab:PPer}
	\end{table*}	

\subsection{Joint Cell Association and Power Control  Schemes}
A very few work in the literature have  considered the problem of distributed CAPC jointly (e.g.,  \cite{yates1995integrated}) with guaranteed convergence. For single-tier networks,  a  distributed framework for uplink was developed \cite{yates1995integrated}, which performs cell selection based on the  effective-interference (ratio of instantaneous interference to channel gain) at the BSs and minimizes the aggregate uplink transmit power while attaining users' desired SIR targets.   Following this approach, a unified distributed algorithm  was designed in \cite{vudistributed} for two-tier networks. The cell association is based on the  \emph{effective-interference} metric and is integrated with a hybrid power control (HPC) scheme which is a combination of TPC and OPC power control algorithms. 

Although the above frameworks are distributed and optimal/suboptimal with guaranteed convergence in conventional networks, they may not be directly compatible to the 5G multi-tier networks. The interference dynamics in multi-tier networks depends significantly on the  channel access protocols (or scheduling), QoS requirements and  priorities at different tiers. Thus, the existing CAPC optimization problems should be modified to include various types of cell selection methods (some examples are provided in Table~I) and power control methods with different objectives and interference constraints (e.g., interference constraints for macro cell UEs, picocell UEs, or D2D receiver UEs). A qualitative comparison among the existing CAPC schemes along with the open research areas are highlighted in Table~II.  A discussion on how these open problems can be addressed  is provided in the next section.

\section{Design Guidelines for Distributed CAPC Schemes in 5G Multi-tier  Networks}	
\label{challenges&suggestions}

Interference management in 5G networks requires efficient distributed  CAPC schemes  such that each user can possibly connect simultaneously to multiple BSs (can be different for uplink and downlink), while achieving load balancing in different cells and guaranteeing interference protection for the HPUEs. In what follows, we provide a number of suggestions to modify the existing schemes.

\subsection{Prioritized Power Control}
To guarantee interference protection for HPUEs, a possible strategy is to modify the existing  power control schemes listed in the first column of Table~II such that the LPUEs  limit their transmit power to keep the interference caused to the  HPUEs below a predefined threshold, while tracking their own objectives. In other words, as long as the HPUEs are protected against existence of LPUEs, the LPUEs could employ an existing distributed power control algorithm to satisfy a predefined goal. This offers some fruitful direction for future research and investigation as stated in Table~II.  To address these open problems in a distributed manner, the existing schemes should be modified so that the LPUEs in addition to setting their transmit power for tracking their objectives, limit their transmit power to keep their interference on receivers of HPUEs below a given threshold. This could be implemented by sending a command from HPUEs to its nearby LPUEs (like a closed-loop power control command used to address the near-far problem), when the interference caused by the LPUEs to the HPUEs exceeds a given threshold. We refer to this type of power control as \emph{prioritized power control}.
Note that the notion of priority  and thus the need of prioritized power control exists implicitly in different scenarios of 5G networks, as briefly discussed in Section~II.
Along this line, some modified power control optimization problems are formulated for 5G multi-tier networks in second column of Table~II.
 
 To compare the performance of existing distributed power control algorithms, let us consider a prioritized multi-tier cellular wireless network where a high-priority tier consisting of $3 \times 3$ macro cells, each of which covers an area of 1000 m $\times$ 1000 m, coexists with a low-priority  tier consisting of $n$ small-cells per each high-priority macro cell, each of which covers an area of 200 m $\times$ 200 m within the coverage area of a high-priority large cell.  Each user is associated with only one BS of its corresponding tier. Each high-priority BS or low-priority BS is located at the centre of its corresponding cell and serves 5 HPUEs and 4 LPUEs, respectively. The target-SIRs are considered to be the same for all users.
 Suppose that all LPUEs employ either TPC, TPC-GR, our proposed prioritized TPC or prioritized TPC-GR, while all of the HPUEs rigidly track their target-SIRs (i.e., HPUEs employ TPC). 

\begin{figure*}[ht]
 \begin{center}
 \includegraphics[width=7 in]{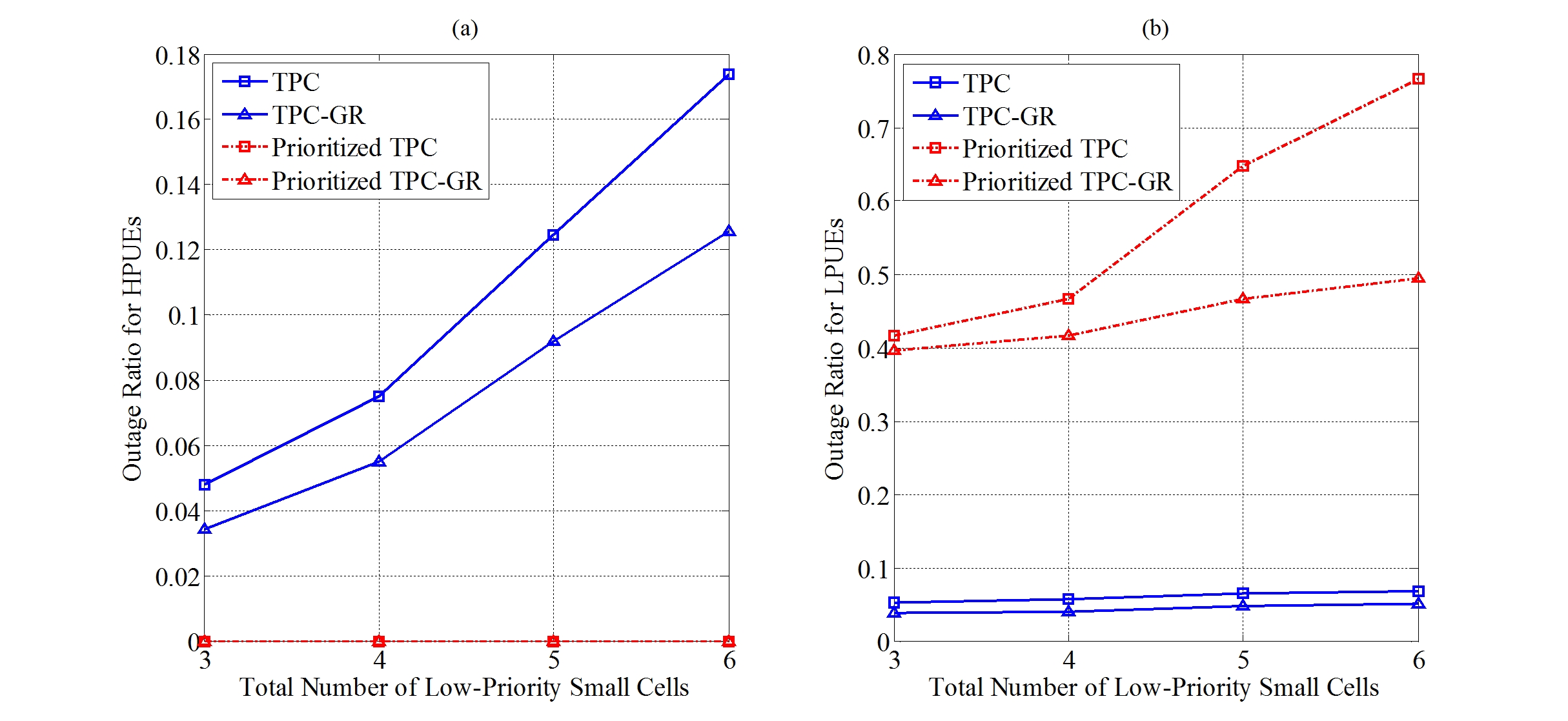}
\hspace{1.2in}(a) \hspace{2.8in}(b)
 \caption [c]{The outage ratio for LPUEs and HPUEs versus the total number of low-priority small cells per each high-priority cell, for the following distributed power control algorithms: TPC \cite{fos93}, TPC-GR \cite{rastisoft}, prioritized TPC, and prioritized TPC-GR.}

 \label{ptpc}
 \end{center}
 \end{figure*}

Fig. \ref{ptpc} illustrates the outage ratio for the HPUEs and LPUEs versus the total number of low-priority small cells (per each high-priority macro cell) ranging from 3 to 6 with step size of 1 (i.e., $n=3, 4, 5, 6$), averaged over 100 independent snapshots for a uniform distribution of BSs and users' locations as explained above. Fig. \ref{ptpc}(a) illustrates the impact of employing either of existing distributed algorithms TPC and TPC-GR employed by LPUEs on the outage of HPUEs (which employ TPC).
 As can be seen, although the outage ratio for the LPUEs and HPUEs are improved by TPC-GR, as compared to TPC, protection of the HPUEs  is not guaranteed by the TPC and TPC-GR algorithms.  In contrast, our proposed prioritized TPC and TPC-GR algorithms guarantee protection of  the HPUEs at the cost of increased outage ratio for the LPUEs. Similar results are achieved when the HPUEs rigidly track their target-SIR by employing TPC and the LPUEs employ the prioritized OPC. That is, protection of the HPUEs is guaranteed at the cost of decreased system throughput for the LPUEs. 
 
 \subsection{Resource-Aware Cell Association Schemes}
Cell association schemes need to be devised that can balance the traffic load  as well as minimize interference or maximize SIR levels at the same time and can achieve a good balance between these objectives without the need of static biasing-based CRE or ABS schemes. As an example, instead of sacrificing the resources of a high-power BS to protect the offloaded users (e.g., as in CRE and ABS technique as detailed in Section IV.A),  user association schemes can also be developed in which a user always prefers to associate with a low-power BS (with no bias) as long as the received interference from high-power BS remains below a threshold. The high-power BS may consider minimizing its transmit power  subject to a maximum interference level experienced by the off-loaded users (i.e., prioritized power control in the downlink).

The CRE technique  forces the users to select  low power nodes by adding a fixed bias to them for traffic load balancing. However, this strategy is  immune to  the resource allocation criterion employed in the corresponding cell. For instance, if a low-power BS performs greedy scheduling, it is highly unlikely that an off-loaded user will get a channel (i.e., low channel access probability) even if the RSRP with bias is the best towards that BS among all other BSs. For round-robin scheduling, if the  low-power BS has a large number of users, it may keep the off-loaded users  in starvation for long time and therefore cause delay.  Clearly, the channel access probability plays a major role in cell-association methods. Thus, the bias selection should  be adaptive (instead of static) to the resource allocation criterion,  traffic load, and distance/channel corresponding to the different BSs. 

In this context,  new cell association schemes/metrics need to be developed
that can optimize multiple objectives, e.g., traffic-load balancing and rate-maximization at the same time.
To illustrate this, we introduce a new \emph{resource-aware} cell association criterion in which each user selects a BS  with {\em maximum channel access probability}, i.e., $\mathrm{max}\{p_i\}$, where $p_i$ is the channel access probability of a cell $i$. Note that, the metric $p_i$ varies for different resource allocation criteria at the BSs. For instance, in round-robin scheduling, $p_i$ is  the reciprocal of the number of users. On the other hand, for greedy scheduling $p_i$ is the probability that the channel gain of a potential admitting user exceeds the channel gain of all existing users in cell $i$ and thus depends on both channel and number of users in cell $i$.
This new metric implicitly tends to balance the traffic load since if the number of users grows in a cell, $p_i$ reduces and stops any further associations or vice versa. In this way, the proposed criterion $p_i$ provides an adaptive biasing to different BSs considering their corresponding scheduling scheme, traffic load 
and channel gains (if opportunistic scheduling is employed).

Note that, in distance-aware cell association, each user selects a cell with minimum distance which tends to improve the sum-rate performance. However, this criterion is immune to traffic load conditions.
Combining the aforementioned resource-aware and distance-aware  criteria,  we now consider a  hybrid cell association.  The hybrid cell association scheme allows a typical user to select a cell with the maximum of product of distance-based channel gain and $p_i$. If $p_i=0$ (i.e., high/infinite traffic load), a user will not select cell $i$ even if its  the closest cell and vice versa. Thus, hybrid schemes assist in achieving  a good balance  between traffic-load balancing   and throughput maximization. 

\begin{figure}[ht]
\begin{center}
\includegraphics[width=4 in]{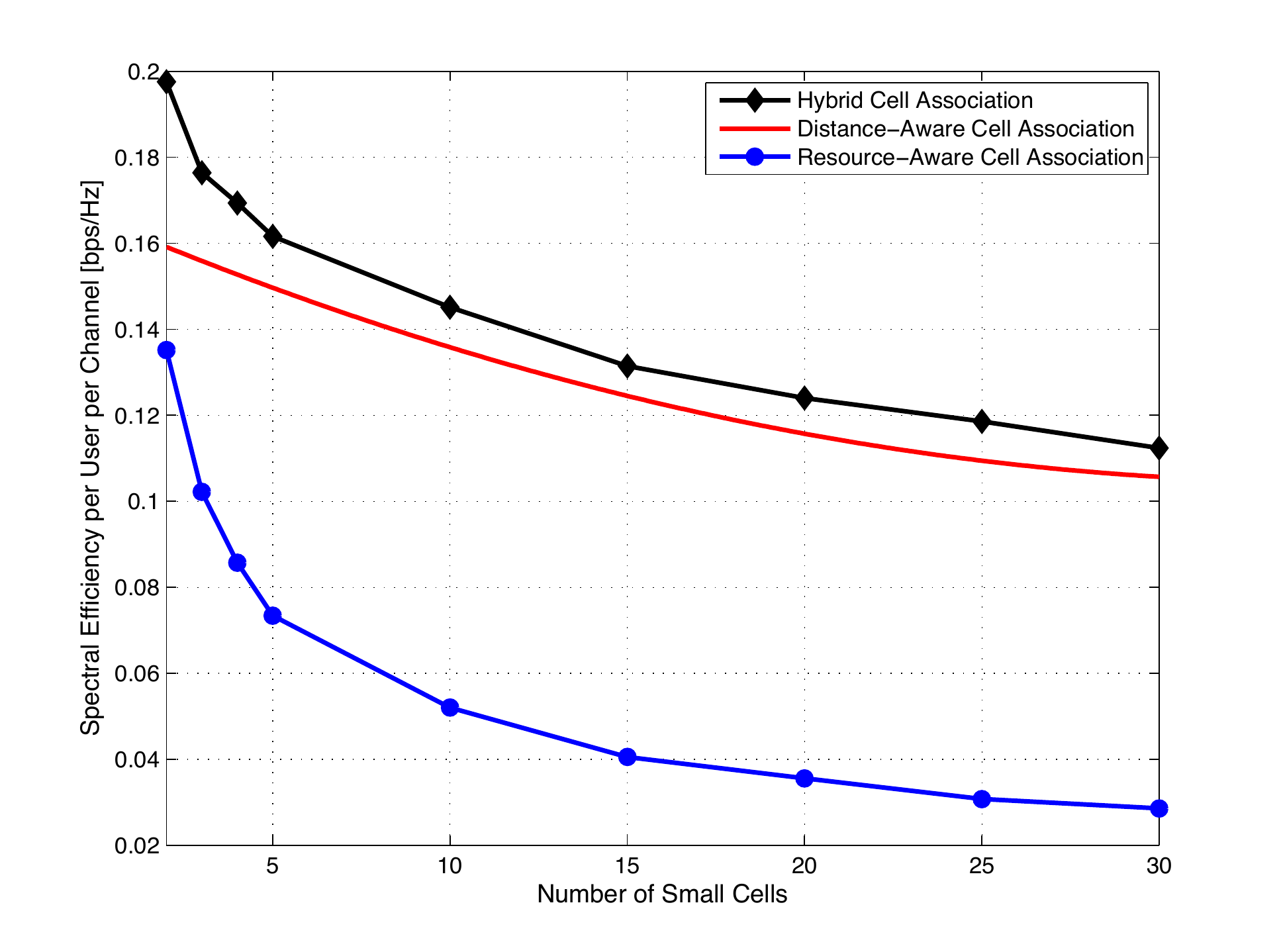}
\caption [c]{Comparison among distance-aware, resource-aware, and hybrid  cell-association schemes (for path-loss exponent $=4$, macrocell   transmit power $=10$~W and small cell transmit power $=1$~W for the simulation setup for a given macrocell shown in Fig.~\ref{systemmodel.eps}).} 
\label{association1}
\end{center}
\end{figure}

For demonstration purpose, a quantitative comparison of a resource-aware, distance-aware and a hybrid cell association scheme is shown in Fig. \ref{association1} for round-robin scheduling. We consider downlink transmission in a multi-tier network with a circular macrocell overlaid by randomly deployed small cells. The average spectral efficiency of a macro user on a given channel is computed as a function of the number of small-cells in a macrocell. Each small cell has a different user intensity with arrival of users modeled by Poisson distribution, thus, creating a non-uniform traffic load scenario per small cell. It can be seen that only resource-aware association (which reduces to load-aware cell association for round-robin scheduling) can degrade the spectral efficiency performance significantly due to strong close-by interferers. 
On the other hand, a hybrid scheme can achieve significant performance gains per channel for a typical user, especially in sparse deployments. 

\begin{figure}[ht]
\begin{center}
\includegraphics[width=4 in]{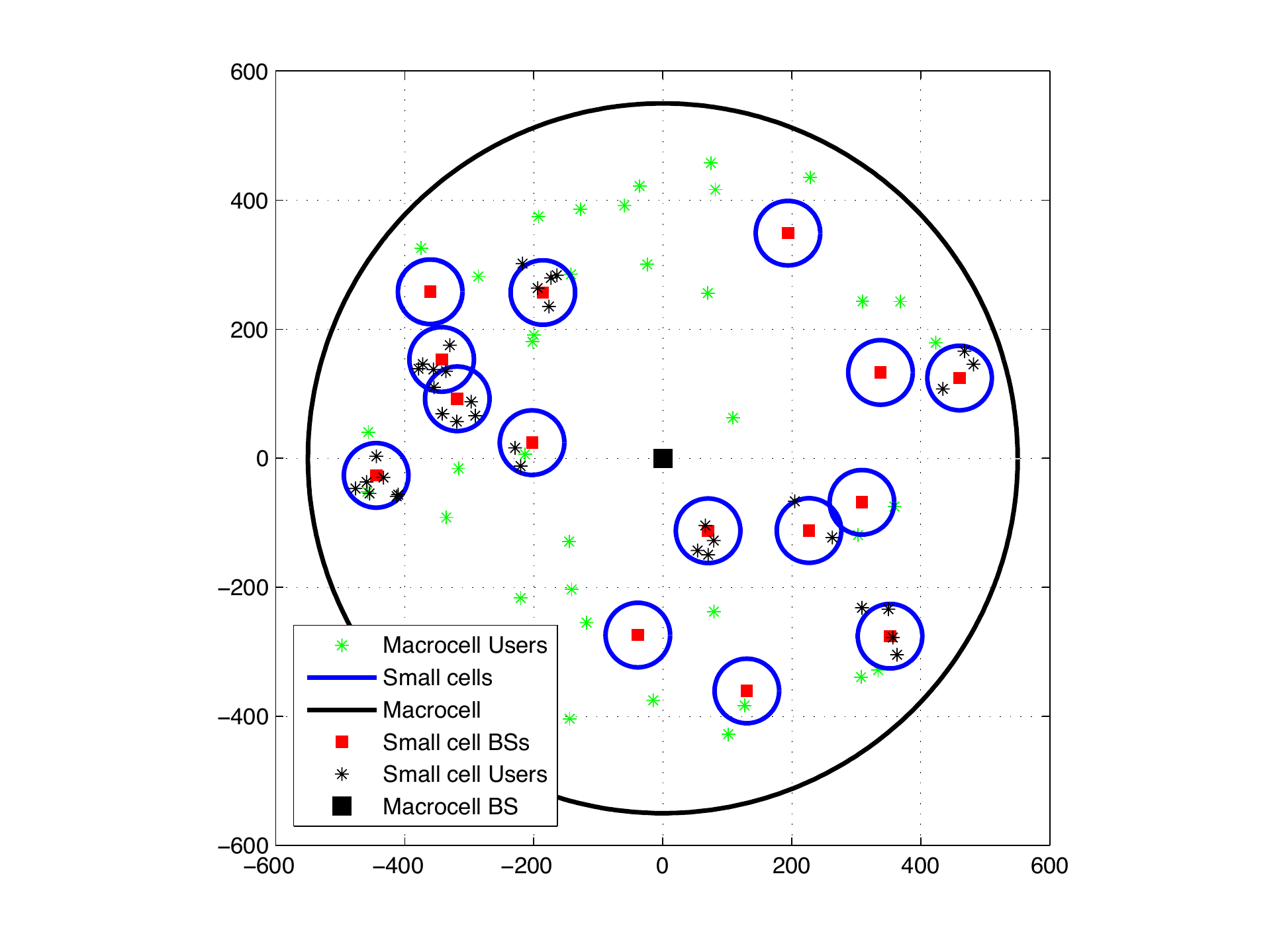}
\caption [c]{A circular macrocell with several small cells. Each small cell has varying user traffic load.}
\label{systemmodel.eps}
\end{center}
\end{figure}

\subsection{Resource-Aware Cell Association and Prioritized Power Control}

Simultaneous connections to multiple BSs and different BS association for uplink and downlink would increase the degrees of freedom which can be exploited to further improve the network capacity and balance the load among different BSs in different tiers. The existing criteria for cell association can be generalized to support simultaneous connection to multiple BSs. For instance, the minimum effective-interference-based cell association can be generalized so that when the differences among effective-interference levels between a given user and some BSs which offer that user the lowest effective-interference levels is not large, that user can simultaneously connects to  those BSs. The proposed resource-aware criterion for cell association can then be combined with this criterion to balance the traffic load.

These cell-association methods can be combined with the prioritized power control schemes depending on the desired objectives. An important issue in this regard is to select a correct combination of cell-association and power control method to achieve a given objective. For instance, joint minimum effective-interference based cell association and OPC is not capable of addressing the  objective of throughput maximization (P3) in the uplink, as in this case all users  try to associate with a BS of minimum effective interference which ultimately results in high transmit power of all users. Although the system throughput is improved when users with good channel conditions increase their transmit power,  it degrades when users with poor channel conditions increase their transmit power. Thus, the need to consider  a different cell-association scheme to achieve the objectives of prioritized power control is evident. In conjunction with OPC, it may be useful to consider RSRP or RSRQ-based cell-association techniques that will allow cell associations based on their channel conditions rather than the received interference. 

\section{Conclusion}
We have outlined the challenges for  interference management in 5G multi-tier networks considering its visions, requirements, and key features. These networks will be characterized by the existence of different access priority for users and tiers along with the  possibility of simultaneous connectivity of users to multiple BSs. Along with these features, different BS association for uplink and downlink transmission open  new challenges and at the same time increase degrees of freedom for power control and cell association. Open challenges have been highlighted and guidelines have been provided to modify the existing schemes in order to make them suitable for 5G multi-tier networks.
In this context, a promising direction for future research is to devise efficient joint CAPC methods that satisfy  objectives such as maximizing system throughput, balance traffic load subject to a minimum SIR for high priority users. To address these multiple objectives, resource-aware user association can be combined with conventional cell association methods to satisfy the required objectives. The hybrid cell association methods combined with prioritized power control will be among  the key enablers for evolving 5G cellular networks.

\bibliographystyle{IEEEtran}

\begin{IEEEbiography}
{Ekram Hossain} (S'98-M'01-SM'06)  
is a Professor in the Department of Electrical and Computer Engineering at University of Manitoba, Winnipeg, Canada. He received his Ph.D. in Electrical Engineering from University of Victoria, Canada, in 2001. Dr. Hossain's current research interests include design, analysis, and optimization of wireless/mobile communications networks, cognitive radio systems, and network economics.  He has authored/edited several books in these areas (http://home.cc.umanitoba.ca/$\sim$hossaina). Dr. Hossain  serves as the Editor-in-Chief for the {\em IEEE Communications Surveys and Tutorials}  and an Editor for {\em IEEE Journal on Selected Areas in Communications - Cognitive Radio Series} and {\em IEEE Wireless Communications}.  Also, he is a member of the IEEE Press Editorial Board. Previously, he served as the Area Editor for the {\em IEEE Transactions on Wireless Communications} in the area of  ``Resource Management and Multiple Access'' from 2009-2011 and an Editor for the {\em IEEE Transactions on Mobile Computing} from 2007-2012. He is also a member of the IEEE Press Editorial Board. Dr. Hossain has won several research awards including the University of Manitoba Merit Award in 2010 (for Research and Scholarly Activities), the 2011 IEEE Communications Society Fred Ellersick Prize Paper Award, and the IEEE Wireless Communications and Networking Conference 2012 (WCNC'12) Best Paper Award. He is a Distinguished Lecturer of the IEEE Communications Society (2012-2015). Dr. Hossain is a registered Professional Engineer in the province of Manitoba, Canada. 
\end{IEEEbiography}

\begin{IEEEbiography}
{Mehdi Rasti} (S'08-M'11) received his B.Sc. degree from Shiraz University, Shiraz, Iran, and the M.Sc. and Ph.D. degrees both from Tarbiat Modares University, Tehran, Iran, all in Electrical Engineering in 2001, 2003 and 2009, respectively. From November 2007 to November 2008, he was a visiting researcher at the Wireless@KTH, Royal Institute of Technology, Stockholm, Sweden. From September 2010 to July 2012 he was with Shiraz University of Technology, Shiraz, Iran, after when he joined the Department of Computer Engineering and Information Technology, Amirkabir University of Technology, Tehran, Iran, where he is an Assistant Professor. From June 2013 to December 2013, he was a postdoctoral researcher in the Department of Electrical and Computer Engineering, University of Manitoba, Winnipeg, MB, Canada. His current research interests include radio resource allocation in wireless networks and network security.
\end{IEEEbiography}

\begin{IEEEbiography}
{Hina Tabassum} received the B.Eng. degree in electronic engineering from the N.E.D University of Engineering and Technology (NEDUET), Karachi, Pakistan, in 2004. During her undergraduate studies she received the Gold medal from NEDUET and from SIEMENS for securing the first position among all engineering universities of Karachi. She then worked as a lecturer in NEDUET for two years. In September 2005, she joined the Pakistan Space and Upper Atmosphere Research Commission (SUPARCO), Karachi, Pakistan and there she received the best performance award in 2009. She completed her Masters and Ph.D. degrees in communications engineering, respectively, from NEDUET in 2009 and King Abdullah University of Science and Technology (KAUST), Makkah Province, Saudi Arabia, in May 2013. Currently, she is working as a post-doctoral fellow in the University of Manitoba, Canada. Her research interests include wireless communications with focus on interference modeling, spectrum allocation, and power control in heterogeneous networks.
\end{IEEEbiography}

\begin{IEEEbiography} 
{Amr Abdelnasser} (S'12) received his B.Sc. and M.Sc. degrees both in Electrical Engineering from Ain Shams University, Cairo, Egypt, in 2006 and 2011, respectively. Currently, he is a Ph.D. student in the Department of Electrical and Computer Engineering, University of Manitoba, Canada. His current research interests include interference management and resource allocation in heterogeneous cellular networks. He has served as a reviewer for several major IEEE conferences and journals. 
\end{IEEEbiography}

\end{document}